\documentclass[journal,hideappendix]{vgtc}                     

\onlineid{1146}

\vgtccategory{Research}

\vgtcpapertype{technique}

\title{IntuiTF: MLLM-Guided Transfer Function Optimization for Direct Volume Rendering}

\author{Yiyao Wang, Bo Pan, Ke Wang, Han Liu, Jinyuan Mao, Yuxin Liu, Minfeng Zhu, Xiuqi Huang, Weifeng Chen,  \\ Bo Zhang, Wei Chen}

\authorfooter{
  \item
    Yiyao Wang, Bo Pan, Ke Wang, Han Liu, Jinyuan Mao, Yuxin Liu, Xiuqi Huang and
    Wei Chen are with the State Key Lab of CAD\&CG, Zhejiang University, and
    Wei Chen is also with the Laboratory of Art and Archaeology Image
    (Zhejiang University), Ministry of Education, China. E-mail:
    \{wangyiyao\, $|$ bopan\, $|$ wangke088\, $|$ 3230103768\, $|$ jymao\, $|$ yuxin.liu\, $|$ chenvis\}@zju.edu.cn.
  \item Weifeng Chen is with Zhejiang University of Finance\&Economics. E-mail: cwf818@zufe.edu.cn.
  \item Minfeng Zhu, Bo Zhang is with Zhejiang University. E-mail: \{minfeng\_zhu@zju.edu.cn, $|$ zhangboknight@gmail.com\}.
}

\abstract{
Direct volume rendering (DVR) is a fundamental technique for visualizing volumetric data, where transfer functions (TFs) play a crucial role in extracting meaningful structures. However, designing effective TFs remains unintuitive due to the semantic gap between user intent and TF parameter space. Although numerous TF optimization methods have been proposed to mitigate this issue, existing approaches still face two major challenges: the vast exploration space and limited generalizability. 
To address these issues, we propose IntuiTF, a novel framework that leverages Multimodal Large Language Models (MLLMs) to guide TF optimization in alignment with user intent. Specifically, our method consists of two key components: \ding{172} an evolution-driven explorer for effective exploration of the TF space, and \ding{173} an MLLM-guided human-aligned evaluator that provides generalizable visual feedback on rendering quality. The explorer and the evaluator together establish an efficient Trial-Insight-Replanning paradigm for TF space exploration. We further extend our framework with an interactive TF design system. We demonstrate the broad applicability of our framework through three case studies and validate the effectiveness of each component through extensive experiments.
We strongly recommend readers check our cases, demo video, and source code at: \url{https://github.com/wyysteelhead/IntuiTF}.

}

\usepackage{caption}
\usepackage[bottom]{footmisc}

\usepackage{upgreek}
\usepackage[dvipsnames,svgnames]{xcolor}
\usepackage{pdfpages}
\usepackage{CJKutf8} 
\usepackage{amssymb}  
\usepackage{pifont}
\usepackage{threeparttable}

\definecolor{bdcolor}{HTML}{d7dce4} 
\definecolor{flcolor}{HTML}{334757} 

\teaser{
  \centering
  \includegraphics[width=\textwidth]{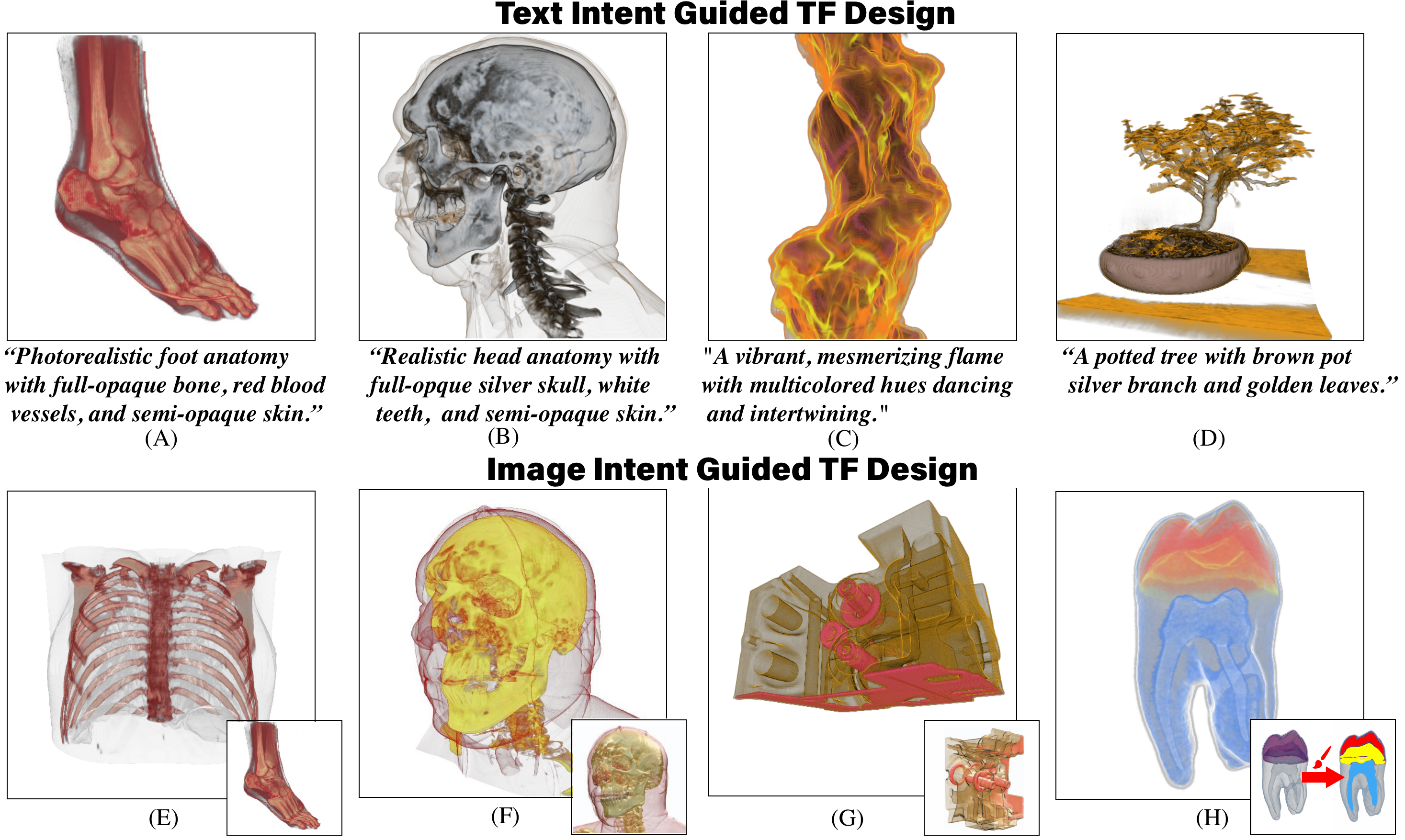}
  \caption{Examples of transfer functions designed using our approach. The first row (A–D) illustrates text‑guided designs, with the corresponding user descriptions shown below each image.
The second row (E–H) illustrates image‑guided designs, where each bottom‑right inset shows the user‑provided reference: (E) reuses the rendered foot from (A), (F) and (G) uses images from other papers, and (H) uses sketch annotations on a rendered tooth to indicate desired coloring.}  \label{fig:teaser}
}

\graphicspath{{figs/}{figures/}{pictures/}{images/}{./}} 

\usepackage{tabu}                      
\usepackage{booktabs}                  
\usepackage{lipsum}                    
\usepackage{mwe}                       
\usepackage{float}
\usepackage{caption}
\usepackage[noend]{algpseudocode}
\usepackage{algorithmicx,algorithm}
\usepackage{graphicx}
\usepackage{makecell} 
\usepackage{booktabs}
\newlength\myindent
\usepackage{mathptmx}                  
\usepackage{amsmath}

\begin{document}


\maketitle
\section{Introduction} 

Direct volume rendering (DVR) is a fundamental technique for visualizing volumetric data in scientific visualization. Transfer function (TF) plays a crucial role in mapping volumetric data to optical properties such as color and opacity \cite{ljung2016state}. A well-designed TF not only enhances image fidelity but also underpins downstream interactions and analysis~\cite{tang2025texgs, ai2025nli4volvis}. Nevertheless, designing appropriate TFs remains a longstanding challenge \cite{Pfister2001bakeoff}. TF design is inherently an iterative process: experts systematically trial different candidates by exploring the TF space, gain insights through visual inspection of these candidates, and replan their search strategy to better align with their analytical goals. This reliance on a manual, iterative loop makes TF design a tedious, trial-and-error process that demands significant domain expertise.

Over the past two decades, researchers have explored numerous methods for intuitive TF design. Early approaches attempted to facilitate TF design via optimization guided by quantitative measures \cite{correa2011visibility, correa2009visibility, wu2010quan}. However, these approaches sacrifice reliable visual assessment by substituting human judgment with static heuristics. Moreover, these heuristic-based metrics often fail to align with diverse analytical objectives, as they optimize for predefined mathematical properties rather than user-specific goals.

Another direction attempted to leverage domain expertise through machine learning \cite{ai2025nli4volvis, Pan2023differentiable, berger_generative_2019}. These approaches have achieved more intuitive TF design by incorporating established domain priors. However, such priors constrain TF exploration from the full parameter space down to predefined knowledge. Additionally, expert knowledge varies significantly across different volume types, making these priors inherently volume-specific. These dual limitations fundamentally restrict their ability to align with novel analytical objectives beyond their training scope.

A third line of research has focused on simplifying the TF design process by enabling more intuitive user specification of analytical goals. Several methods allow users to specify their intentions through high-level inputs such as text or images, with the system handling the translation to TF parameters \cite{jeong2024text, liu2024ava}. Recent explorations have integrated multimodal large language models (MLLMs) with visualization platforms to enable more natural interaction with visualization tools \cite{liu2025paraview}. While these approaches improve initial alignment between user intent and TF parameters, they frame TF design primarily as user‑driven specification, thereby placing the burden of exploration entirely on users.

Despite these advances, current approaches still fail to fully address the fundamental challenge of intuitive TF design. Despite their diverse strategies, current methods share a common approach: they attempt to replace or bypass the expert's iterative exploration process. We believe that an intuitive TF design framework should not eliminate this process, but instead study and formalize the underlying exploration patterns.

In this paper, we propose IntuiTF (\underline{Intui}tive \underline{T}ransfer \underline{F}unction Design), a framework that mimics expert-driven exploration by modeling the expert's TF design process.
Our core insight is to formulate the expert's transfer function design process into a three-stage cycle: Trial-Insight-Replanning. This formalization captures how experts systematically generate diverse candidates, evaluate their quality and alignment with analytical goals, and strategically refine their search direction.
IntuiTF realizes this formalization through a transfer function optimization approach comprising two core components: an evolution-driven explorer and a human-aligned evaluator. The explorer manages the Trial and Replanning phases using an evolutionary algorithm that encodes transfer functions into efficient genetic representations, generates solutions through mutation and crossover operations, and selects high-quality, diverse candidates for subsequent iterations. The evaluator employs a specialized MLLM-guided direct volume rendering assessment method that approximates human visual assessment, ranking candidate solutions based on both visualization quality and user intent alignment.
We further implement an interactive transfer function design system to demonstrate diverse applications through three case studies, and validate the effectiveness of each component through extensive quantitative and expert interviews.

In summary, our primary contributions include:
\begin{itemize}[leftmargin=12pt, itemsep=0pt, topsep=0mm]
    \item A transfer function design framework that models and implements the expert's iterative 'Trial-Insight-Replanning' workflow.
    
    
    \item An MLLM-guided evaluator for DVR quality assessment that provides human-aligned feedback to guide TF design process.
    
    \item A systematic analysis of diverse application cases demonstrating the effectiveness and adaptability of our approach across different visualization scenarios.
    
\end{itemize}

\section{Related Work}

\begin{figure*}[t!]
    \centering
    \includegraphics[width=\textwidth]{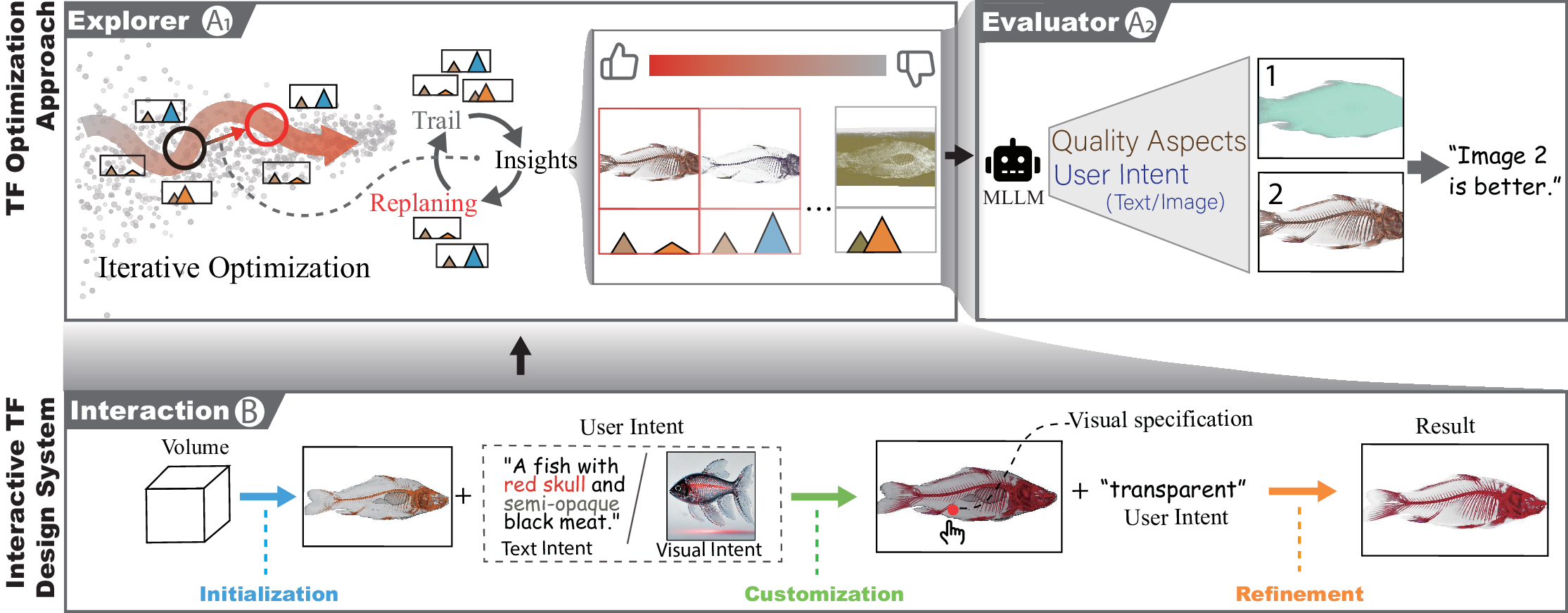}
    \caption{Overview of our TF design framework IntuiTF. (A) Our transfer function optimization approach combines: ($\text{A}_1$) an evolution-driven explorer following trial, Insights, and Replaning optimization process formulation that iteratively generates and refines TFs with replanning capabilities, and ($\text{A}_2$) an MLLM-guided evaluator that assesses visualizations based on formal quality aspects and user intent. (B) Our interactive transfer function design system includes: a three-stage interaction process of initialization-customization-refinement, which enables users to express visualization intentions through text descriptions or reference graphs and gradually refine the results.}
    \vspace{-1em}
    \label{fig:workflow}
\end{figure*}

\textbf{Optimization-based Transfer Function Design.}
Early research has formalized transfer function design as a parameter optimization problem \cite{HHKP96}. By parameterizing transfer functions as piecewise linear functions \cite{kniss2002multidimensional} or Gaussian mixture models \cite{kniss2003Gaussian}, the problem was transformed into a search task in high-dimensional parameter space through automatic or semi-automatic optimization methods. Global optimization methods such as evolutionary algorithms have been widely applied due to their capability in handling TF design. He et al. \cite{HHKP96} implemented multiple stochastic algorithms including genetic algorithms, simulated annealing, and hill climbing for transfer function optimization. Wu et al. \cite{wu_interactive_2007} used genetic algorithms to automatically generate transfer functions matching user intent.

These methods primarily rely on optimization guidance from two types of quantitative metrics: statistical features of volume data \cite{tzeng2005intelligent,maciejewski_structuring_2009,5429612,Wang:2011:EVEUGMM,LLY06a,Kindlmann:1999:SAG, KUS∗05} or visual properties of rendered results \cite{correa2011visibility,correa_size-based_2008,luo2018intuitive,shiaofen_fang_image-based_1998,marks_design_1997,PF08,prauchner_two-level_2005,hanqi_guo_transfer_2014,JFY15,guo2011WYSIWYG,HHKP96}. However, the quality ceiling of these methods is fundamentally constrained by their analytical capabilities. Quantitative analysis cannot effectively simulate human visual analysis processes, resulting in significant gaps between optimization results based on these metrics and expert expectations. While user interaction can provide some degree of guidance, such interaction is itself limited by the expressive power of underlying metrics and cannot fundamentally solve the problem. 

Building on this foundation, our work leverages evolutionary algorithms for systematic exploration while replacing traditional quantitative metrics with MLLM-guided visual assessment specialized for DVR scenario.

\textbf{Domain Prior-based Transfer Function Design.}
To avoid requiring users to perform complex parameter space exploration, learning-based methods take on the exploration task by equipping models with prior design knowledge \cite{ai2025nli4volvis, tang2025texgs, Pan2023differentiable, berger_generative_2019, nguyen2021finding, tang2024stylerf,tang2025ivr,yang2025meta}. Using various machine learning techniques, these approaches train model on volume data paired with high-quality transfer functions, learning mappings from data features to transfer function parameters. Pan et al. leveraged neural rendering to create latent design spaces for transfer function exploration \cite{Pan2023differentiable}. Recent advances have explored alternative representations that move beyond traditional transfer function paradigms. NLI4VolVis \cite{ai2025nli4volvis} extract and understand semantic components in volume renderings by turning volume into editable 3D Gaussian splatting representations. TexGS-VolVis \cite{tang2025texgs} trains its volume scene representation based on multiple basic TFs that are manually designed.

However, the performance of these methods heavily depend on the quality and coverage of training data. Due to the scarcity and scattered distribution of volumetric datasets, existing approaches can only generalize expert experience within individual volume data or limited data domains. When faced with out-of-distribution data or novel analysis requirements, the generalization capability of models becomes insufficient, resulting in unsatisfactory outcomes. The dependence on static prior knowledge fundamentally limits their application potential in open-ended exploration scenarios. 

Unlike these approaches, our method take advantage of the general visual reasoning capabilities of pre-trained MLLMs, enabling our method to generalize across volume datasets and analytical objectives without domain-specific training.

\textbf{Intent-driven Transfer Function Design.}
Another line of research focuses on simplifying the interaction cost of user exploration, enabling users to express design intent in more intuitive ways. These methods allow users to specify analysis objectives through high-level inputs such as natural language descriptions or example images \cite{jeong2024text,liu2025paraview, liu2024ava,mallick2024chatvis}, with the system responsible for converting these intuitive inputs into underlying transfer function parameters. Recent work by Jeong et al. presented a text-to-transfer function network to minimize design complexity \cite{jeong2024text}. The development of multimodal large language models has further advanced the realization of such intuitive interaction approaches, with systems like ParaView-MCP \cite{liu2025paraview} demonstrating how MLLMs can enable natural language and visual interaction with visualization tools through standardized interfaces.


Although these methods lower interaction barriers, they still treat transfer function design as user-driven parameter specification. When facing ambiguous intent or unknown datasets, systems lack autonomous exploration capabilities, leaving cognitive burden entirely to users. Our approach combines intuitive intent specification with autonomous exploration, enabling systematic generation and evaluation of transfer function candidates that align with analytical goals while reducing manual exploration burden.

\section{IntuiTF Overview}\label{sec:overview}


Fig.~\ref{fig:workflow} presents an overview of IntuiTF. We develop a novel framework that simulates the TF design process of human visualization experts. By implementing this framework, we enable users to gain insights based on automatically explored transfer functions, and express their design intentions through intuitive interactions.

Fig.~\ref{fig:workflow}(A) illustrates our transfer function optimization approach's architecture, which comprises two principal modules. The Explorer module (Fig.~\ref{fig:workflow}, $\text{A}_1$) iteratively explores the TF parameter space to identify optimal solutions that fulfill the visualization goals. Inspired by how experts design TFs, our Explorer begins by exploring multiple trials from the current design state, making incremental adjustments to sample new possibilities in the design space. It then analyzes visual feedback from these trials to gain insights about which adjustments lead to improved results. Based on these insights, it dynamically replans its search trajectory, continuously refining its strategy through this iterative trial-replanning cycle. We implement this using a progressive evolutionary algorithm that efficiently traverses the high-dimensional parameter space while avoiding local optima. The Evaluator module (Fig.~\ref{fig:workflow}, $\text{A}_2$) provides critical visual feedback to guide the Explorer's search process. It mimics how experts would assess visualization quality through evaluation aspects that consider both formal quality assessments and user-specified intents. The Evaluator's ability to align with both human visual perception standards and specific user intentions is crucial for producing high-quality results that match user expectations.

Leveraging this transfer function optimization approach, we design a interactive transfer function design system that enables intuitive TF design. Fig.~\ref{fig:workflow} (B) illustrates our system's three-stage interaction. The Initialization stage automatically generates diverse high-quality visualizations to establish a starting point (e.g., revealing major structures such as the skull and flesh of a fish). In the Customization stage, users can express their visualization preferences through dual modalities: textual intent using natural language descriptions or visual intent by providing reference images. The system then adapts the TF parameters to match these high-level intentions. During the Refinement stage, users can make targeted adjustments to specific features without disrupting the overall visual composition (e.g., clicking on the flesh and specifying “transparent” to better expose the skeleton).

In the following sections, we elaborate on the details of our framework, IntuiTF. Sec. \ref{sec:approach} presents our formulation of the expert transfer function design process and introduces our Explorer-Evaluator approach, designed to leverage MLLMs for improving this process. Sec. \ref{sec:interaction} details our interactive transfer function design system, including both the user interface and the interaction implementation. Sec. \ref{sec:case_study} illustrates the application of IntuiTF in various visualization scenarios through three case studies. Sec. \ref{sec:experiment} presents quantitative experiments that validate the efficiency of our approach. Finally, Sec. \ref{sec:expert} reports on expert feedback through interviews.

\section{Transfer Function Optimization Approach}\label{sec:approach}

\subsection{TF Design Process Formulation}\label{sec:formulation}


To automate the transfer function design process, we first formalize human behavior during transfer function design. 
We begin by defining the fundamental elements of the transfer function design process. Let \( D \) denote the volumetric dataset, \( T \) represent the transfer function design space, and \( V \) be the visualization result space. A transfer function \( t \in T \) maps volume data values to optical properties, and the volume rendering process is formulated as: $R: (D, T) \rightarrow V$. Transfer function design is an iterative optimization process that seeks an optimal transfer function \( t^* \in T \), such that the corresponding visualization result $v^* = R(D, t^*)$ best satisfies the user's visual requirements.

Inspired by research on the transfer function design process \cite{Pfister2001bakeoff}, we decompose the human transfer function design process into three key optimization steps: Trial, Insight, and Replanning.

\textbf{Trial.} 
During the Trial step, designers make multiple small parameter adjustments to the current transfer function $t_0$, generating a set of perturbed functions $\{\tilde{t}_k\}_{k=1}^n$. These small adjustments help maintain existing visual features while gradually observing the visual changes, preventing the loss of valuable information.

\textbf{Insight.} 
In the Insight step, designers analyze the visualization results corresponding to the perturbed transfer functions using a set of evaluation metrics $\mathcal{M} = \{\mu_1, \mu_2, ..., \mu_m\}$. These metrics encompass multiple dimensions of visual quality assessment. By examining the relationships between parameter adjustments and visual outcomes, designers gain insights $I = \phi(\{(\tilde{t}_k, \tilde{v}_k, \mu(\tilde{v}_k, v))\}_{k=1}^n)$ about promising directions for further optimization.

\textbf{Replanning.} 
Based on the acquired insights $I$, designers select potentially promising transfer functions from the trial results as starting points for the next iteration and formulate new parameter adjustment strategies accordingly. This Trial-Insight-Replanning cycle continues iteratively until a satisfactory visualization result is achieved.


Based on this formulation, we now need to select an optimization approach that can effectively implement the Trial-Insight-Replanning cycle. After examining various algorithmic alternatives, we have identified an evolutionary algorithm due to its natural correspondence with the Trial-Insight-Replanning cycle.
The structural parallels between evolutionary algorithms and the human design process are remarkably strong \cite{dietrich2015human}. 
First, mutation and crossover operations embody the Trial phase by generating diverse solutions. Second, fitness evaluation implements the Insight phase by assessing solutions against multiple criteria. Finally, selection executes the Replanning phase by retaining fitter individuals for the next generation.

Leveraging the evolutionary algorithm, we develop a system architecture with two main modules: the Explorer (Fig. \ref{fig:workflow}, $\text{A}_1$) implements the Trial and Replanning phases through evolutionary operations customized for the transfer function design task; within this process, the Evaluator (Fig.~\ref{fig:workflow}, $\text{A}_2$) performs the Insight phase through comprehensive quality assessment to inform selection decisions.

\subsection{Evolution-Driven Explorer}
\label{subsec:evolution}

\begin{figure}
    \centering
    \includegraphics[width=0.5\textwidth]{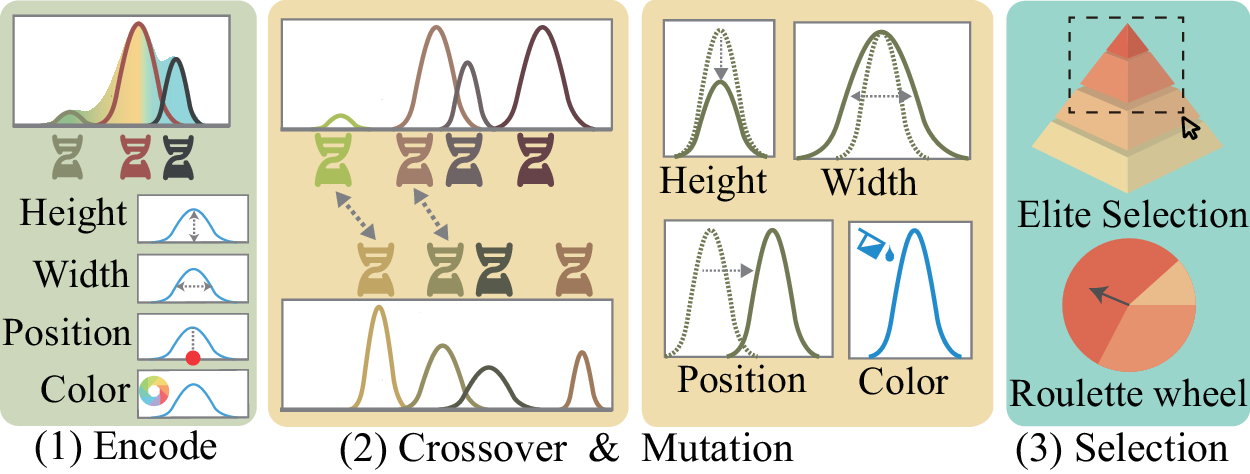}
    \caption{Key components of our evolution-driven explorer. (1) Encoding: Transfer functions are represented as combinations of Gaussian primitives with parameters for height, width, position, and color. (2) Genetic operations: Customized crossover and mutation strategies enable effective exploration of the parameter space. (3) Selection: A roulette wheel selection mechanism biases the evolutionary process toward higher-fitness solutions, strategically guiding the search toward optimal transfer functions.}
    \label{fig:genetic}
    \vspace{-1em}
\end{figure}

Fig.~\ref{fig:genetic} illustrates the key components of our evolution-driven explorer. Our implementation addresses three critical aspects: (1) an efficient genetic encoding of transfer functions that facilitates design space exploration, (2) the mutation and crossover operations that implement the Trial phase by generating diverse candidate solutions, and (3) the selection operation that embodies the Replanning phase by preserving promising directions. Together, these components create an evolutionary method specifically tailored to the transfer function optimization process.

\textbf{Transfer Function Representation.} To parameterize transfer functions for evolutionary operations, we encode them using a parametric Gaussian mixture model (GMM), which offers both smooth transitions that align with human perception and efficient expressiveness with minimal parameters.
Formally, we define our genetic representation as follows: each transfer function is encoded as a genome $G = \{g_1, g_2, ..., g_n\}$, where $n$ represents the number of Gaussian components in the mixture model. A larger value of $n$ enables finer feature discrimination within the volume data, as each Gaussian component typically corresponds to a distinct structural feature in the visualization. Each gene $g_i$ encapsulates a complete parameter set for a single Gaussian component, specifically: $g_i = \{\mu_i, \sigma_i, w_i, c_i\}$, where $\mu_i$ represents the mean (position in data value range), $\sigma_i$ the variance (width of influence), $w_i$ the weight (contribution strength), and $c_i$ the RGB color vector associated with this component.
In our GMM-based transfer function, the opacity at any data value is calculated as the weighted sum of all Gaussian contributions at that position. The color at each data value is determined through linear interpolation between the color parameters of contributing Gaussian components.

\textbf{Mutation and Crossover.} With the genetic representation established, we now focus on the Trial phase implementation through mutation and crossover operations. For crossover, we sequentially examine each Gaussian gene in a pair of parent solutions and exchange their attributes with a predetermined probability. This exchange can involve either a single attribute (color, position, Gaussian shape) or multiple attributes simultaneously, allowing for flexible genetic recombination.
For mutation, each gene has a probability of undergoing one or more of the following transformations: (1) Height mutation modifies the weight parameter $w_i$; (2) Width mutation adjusts the variance $\sigma_i$; (3) Position mutation shifts the mean $\mu_i$; and (4) Color mutation alters the RGB vector $c_i$, with occasional recombination of the component's own color channels. These mutation operations directly mirror the manual adjustments designers typically perform when creating transfer functions: vertical dragging to modify opacity contribution, horizontal dragging to reposition features, width adjustment to control feature boundaries, and color selection to enhance feature discrimination.

\textbf{Selection.} We implement the selection process through a combination of elitism and roulette wheel selection. Elitism ensures that the best-performing individuals from the current generation are preserved and passed on to the next generation, while roulette wheel selection introduces diversity by probabilistically selecting individuals based on their fitness scores. The crucial fitness evaluation that informs this selection process is performed by our MLLM-guided human-aligned Evaluator (detailed in Section \ref{subsec:evaluator}), which employs comprehensive assessment of volume renderings to rank candidate solutions according to both general visualization quality criteria and specific user intent alignment.

\subsection{Human-Aligned Evaluator}
\label{subsec:evaluator}

\begin{figure*}[!t]
    \centering
    \includegraphics[width=\textwidth]{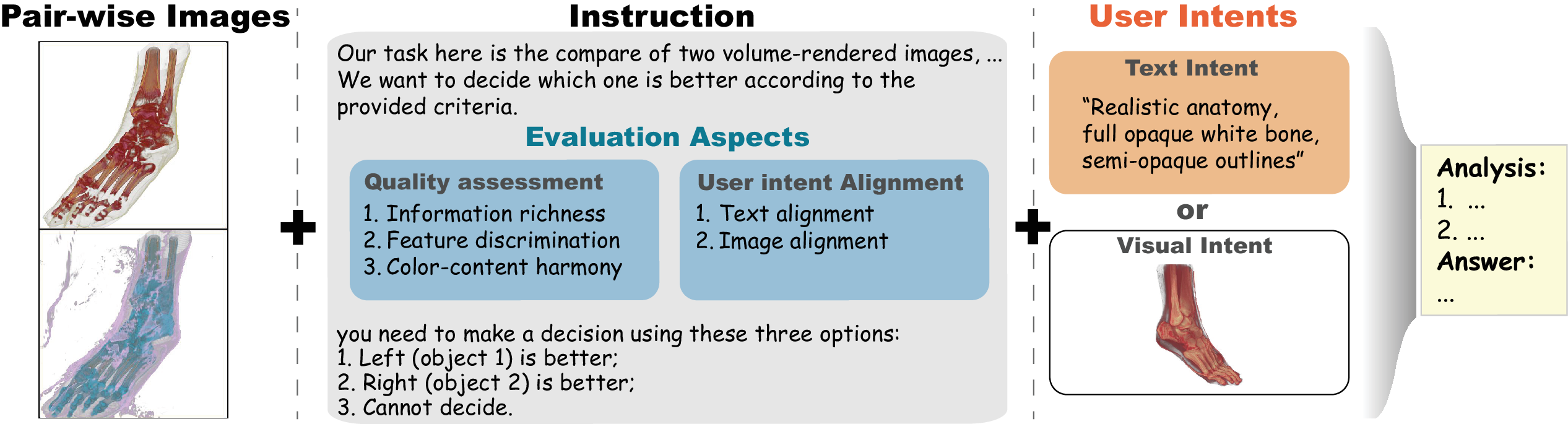}
    \caption{Overview of our MLLM-guided evaluator design. The evaluator takes three key components to make the comparative analysis: (1) A pair of images for the comparative analysis, (2) evaluation aspects that helps MLLM in making comprehensive assessment, and (3) user's intents expressed through text or image.}
    \label{fig:evaluator}
    \vspace{-1em}
\end{figure*}


In the Insight step of transfer function design, human judgment is inherently qualitative rather than quantitative. Designers cannot definitively determine an ``optimal transfer function'' in absolute terms, but instead employ comparative assessments between alternatives. The qualitative nature of human evaluation makes purely quantitative metrics inadequate for assessing transfer function quality. We need an evaluator that can make qualitative assessments that match how humans judge visualization results.

To address this challenge, we turned to recent advances in artificial intelligence. MLLMs have demonstrated remarkable capabilities in image understanding and analysis, with perceptual abilities that align closely with human visual assessment patterns\cite{viseval, chen2024mllm, gpteval3d}. GPTEval3D \cite{gpteval3d} has established that these models can perform human-aligned analysis and evaluation of text-to-3D generations with high correlation to human judgments. The analysis of volume visualizations shares similarities with evaluating 3D rendered content, both involving the perceptual assessment of 3D structural relationships and their 2D visual representations.

We therefore apply MLLMs to our visualization evaluation task, leveraging their demonstrated capabilities in human-aligned volume rendering assessment. As shown in Fig. \ref{fig:evaluator}, our evaluator provides the MLLM with three critical input components: (1) the pair of volume renderings to be compared, (2) evaluation aspects consisting of both formal quality aspects and user intent alignment criteria, and (3) user intent specifications that establish visualization goals through either natural language descriptions or reference images. Our evaluation process works by presenting the two volume renderings to the MLLM and prompting it to analyze them across multiple predefined aspects. The model assesses each evaluation aspect with respect to user intent, and determines which rendering performs better. These aspect-specific judgments are then combined through weighted averaging to derive an overall preference. Through this pairwise comparison process, we can obtain comparative judgments between transfer functions. 

Our evaluator leverages the perceptual capabilities of MLLMs through two key components. First, we design specialized evaluation criteria that guide the model's judgments to align with human perceptual preferences for volume visualization. Second, we develop a computational method to transform qualitative pairwise comparisons into quantitative fitness scores suitable for our evolutionary algorithm. Together, these components enable us to derive meaningful fitness metrics that effectively guide the optimization process.

\textbf{(1) Evaluation aspects.} We summarized three formal quality aspects from previous works in transfer function evaluation \cite{correa2009visibility, ljung2016state}: information richness, feature discrimination, and color harmony. Information richness ensures meaningful features are visible while suppressing noise and irrelevant elements. Feature discrimination enables the extraction of different structural features in the volumetric data. Color harmony ensures visualizations align with human aesthetic preferences. Additionally, our evaluator assesses the alignment with user intent through two key dimensions: text intent and visual intent. For text intent alignment, we instruct MLLMs to evaluate not only textual consistency with user requirements but also the overall visual quality of the rendered results. For visual intent alignment, we first require MLLMs to interpret user-provided reference images in terms of their structural and stylistic features (e.g., as in Fig. \ref{fig:evaluator}, where a foot rendering is interpreted as “bones depicted in realistic reddish-brown tones with semi-transparent soft tissue context”). The assessment then focuses on two crucial aspects: color style matching and structural style matching. Color style matching evaluates how well the visualization adopts the color schemes from reference images. Structural style matching measures how effectively the visualization highlights similar anatomical or pathological structures as shown in reference images.

\textbf{(2) Fitness scores calculation.} To integrate these qualitative comparisons into our evolutionary algorithm, we need a numerical fitness measure. We adopt the Elo rating system \cite{elo1978rating}, widely used in competitive chess and generative model evaluation \cite{gpteval3d}. Let $\sigma_i \in \mathbb{R}$ denote the Elo score of the $i$-th transfer function in our population; the transfer function with higher scores indicates better performance in volume visualizations. To begin the Elo matching process, each transfer function starts with an initial rating of 1600. We conduct a Swiss-system tournament \cite{swiss-system-chessbase} consisting of $\log_2(n) + 2$ rounds, where $n$ is the population size. In each round, transfer functions with similar ratings are paired together while avoiding repeated matchups.

After each pairwise comparison, we calculate the expected outcome $E$ based on current ratings, representing the probability that transfer function $i$ produces a better visualization than transfer function $j$, as shown in Eq. \ref{eq:E}. We then compute the rating adjustment $\Delta$ using the rating adjustment factor $K=32$ and the actual outcome $S$ (1 for win, 0 for loss), as shown in Eq. \ref{eq:delta}. The winner gains $\Delta$ points while the loser loses an equivalent amount, with larger adjustments occurring when results contradict expectations based on current ratings.

\begin{equation}\label{eq:E}
E = \text{Pr}(\text{``}i \text{ better than } j\text{''}) = (\frac{1}{1 + 10^{(\sigma_j-\sigma_i)/400}})^{-1}
\end{equation}

\begin{equation}\label{eq:delta}
\Delta = K \cdot (S - E)
\end{equation}

Ultimately, the Elo rating transforms the evaluator's pairwise comparisons into individual ratings for each transfer function. However, using raw Elo ratings directly as fitness scores for roulette wheel selection would create disproportionate selection probabilities. A critical issue is that Elo rating differences remain relatively constant throughout the evolutionary process, while the actual visualization quality distribution changes significantly. In early iterations, we need broad selection to avoid local optima; in later iterations, we should favor superior results to preserve beneficial characteristics. Raw Elo ratings don't adapt to this changing landscape, potentially hindering optimization efficiency.

Therefore, we convert these Elo ratings into appropriate fitness scores by scaling up the differences as iterations progress. We first translate Elo ratings into rank information, then calculate the fitness for each transfer function as $(n - \text{rank} + 1)^{\text{current\_pressure}}$, where $n$ is the population size. The parameter $\text{current\_pressure}$ controls the selection differential between high and low-ranked transfer functions and varies dynamically with iteration rounds. During early iterations, we maintain a low selection pressure, creating minimal fitness differences between the best and worst transfer functions. As iterations progress, we gradually increase the selection pressure, widening the fitness gap between high and low-ranked functions. We formulate the current pressure as $\text{min\_pressure} + (\text{max\_pressure} - \text{min\_pressure}) \cdot (\text{progress})^k$, where $\text{progress} = \min(1.0, \text{current\_gen} / \text{max\_gen})$. The parameter k controls how quickly this transition occurs, with larger values delaying the pressure increase until later iterations. In our implementation, we set min\_pressure=1.2, max\_pressure=4.0, and k=2.0.

\section{Interactive Transfer Function Design System}\label{sec:interaction}

In this section, we demonstrate how our practical visualization system integrates our transfer function design approach (Fig. ~\ref{fig:workflow}, $\text{A}_1$, $\text{A}_2$) into our system's three-stage interaction (Fig. ~\ref{fig:workflow}, $\text{B}$). We first present the interface design that enables users to effectively communicate their design intents and interact with the generated visualizations. Then, we describe how our approach is configured to support each stage of the interaction.

\subsection{User Interface Details}

\begin{figure}[!t]
    \centering
    \includegraphics[width=0.5\textwidth]{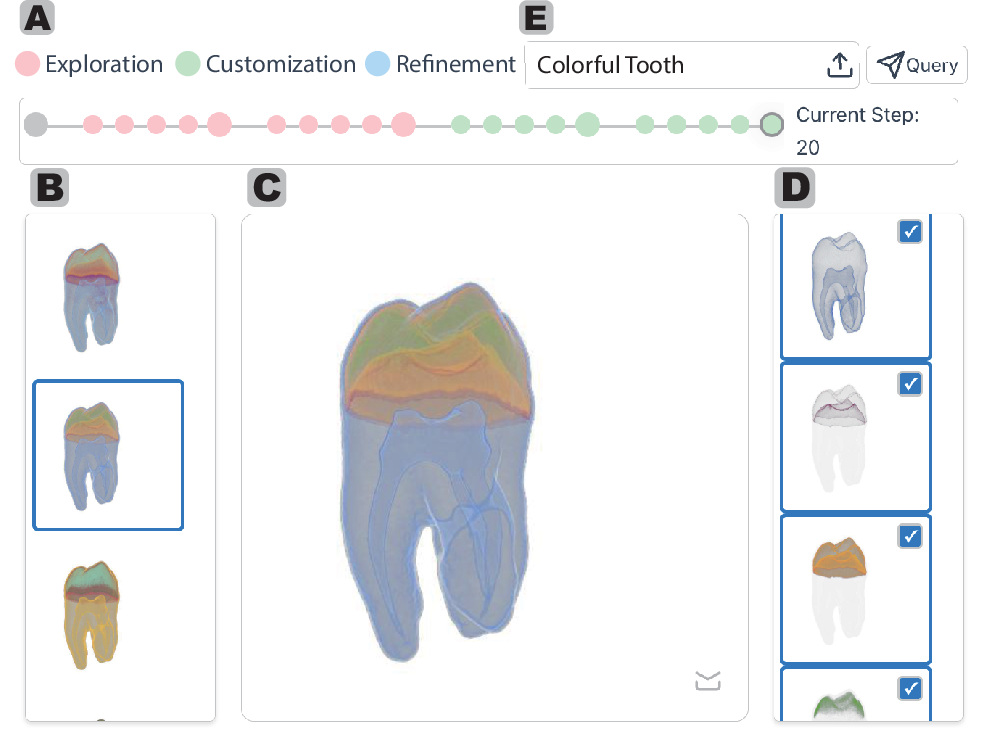}
    \caption{The interface of the interactive transfer function design system.}
    \label{fig:interface}
    \vspace{-1em}
\end{figure}

Fig.\ref{fig:interface} illustrates the user interface of our system, which supports our three-stage interaction paradigm through five integrated components: \raisebox{-0.3em}{\includegraphics[height=1.2em]{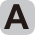}} history view, \raisebox{-0.3em}{\includegraphics[height=1.2em]{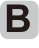}} recommendation gallery view, \raisebox{-0.3em}{\includegraphics[height=1.2em]{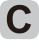}} volume interaction view, \raisebox{-0.3em}{\includegraphics[height=1.2em]{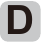}} feature selection view, and \raisebox{-0.3em}{\includegraphics[height=1.2em]{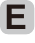}} query view.

The history view \raisebox{-0.3em}{\includegraphics[height=1.2em]{figs/a.pdf}} maintains a record of visualization results generated during the automatic exploration process. This gallery enables users to track the explorer's search and evolution process, as well as return to previous customization points for alternative exploration directions.

The recommendation gallery view \raisebox{-0.3em}{\includegraphics[height=1.2em]{figs/b.pdf}} presents the top-ranked visualization results from the current exploration iteration. Users can browse, compare, and select promising options for further refinement.

The volume interaction view \raisebox{-0.3em}{\includegraphics[height=1.2em]{figs/c.pdf}} functions as the primary visualization workspace, supporting standard rotation, panning, and zooming operations to examine the rendered volume. It also enables direct feature selection through clicking operations that automatically segment and highlight selected components.

The feature selection view \raisebox{-0.3em}{\includegraphics[height=1.2em]{figs/d.pdf}} displays individual volumetric features as separate entities, decomposing complex structures into distinct components. When users select specific features, either through direct clicking in the volume view \raisebox{-0.3em}{\includegraphics[height=1.2em]{figs/c.pdf}} or by choosing from this component, selected features are visually highlighted to confirm user intent and prevent miscommunication.

Finally, the query view \raisebox{-0.3em}{\includegraphics[height=1.2em]{figs/e.pdf}} enables multimodal specification of visualization goals through natural language and visual inputs. Users can articulate preferences textually (``render the skull in semi-transparent red'') or upload reference images demonstrating desired rendering styles. The system interprets both input types alongside selected features to create comprehensive multimodal specifications.

\subsection{Interaction Implementations}

In this chapter, we examine how our technical approach adapts to serve users across the three interaction stages previously outlined in Fig.~\ref{fig:workflow}(B). We implement stage-specific optimizations to address the varying user needs during exploration, customization, and refinement.

\textbf{Initialization}. The goal here is to autonomously generate diverse, high-quality visualizations. To achieve this, the evaluator focuses exclusively on formal quality metrics, while the evolutionary algorithm prioritizes search breadth. We implement relatively high crossover and mutation rates that gradually decrease across iterations. This allows transfer functions to converge toward effective solutions.

\textbf{Customization}. Building on the high-quality visualizations established during exploration, this stage aims to further explore the transfer function space to meet user requirements. Accordingly, crossover and mutation rates are increased again to broaden the search. Meanwhile, mutation rates for the Gaussian position parameter are reduced to preserve the feature structures identified during exploration. 

\textbf{Refinement}. The final stage focuses on enabling fine-grained, local adjustments. To allow users to specify which part should be modified, we enable feature selection through direct interaction with the rendering: when a user clicks on a region of interest, the system identifies all visible features covering that position and and selects the one with the smallest projected area. During subsequent optimization, only the Gaussian components of selected features continue to evolve, while those corresponding to other features remain frozen. 

\begin{figure}[!t]
    \centering
    \includegraphics[width=0.5\textwidth]{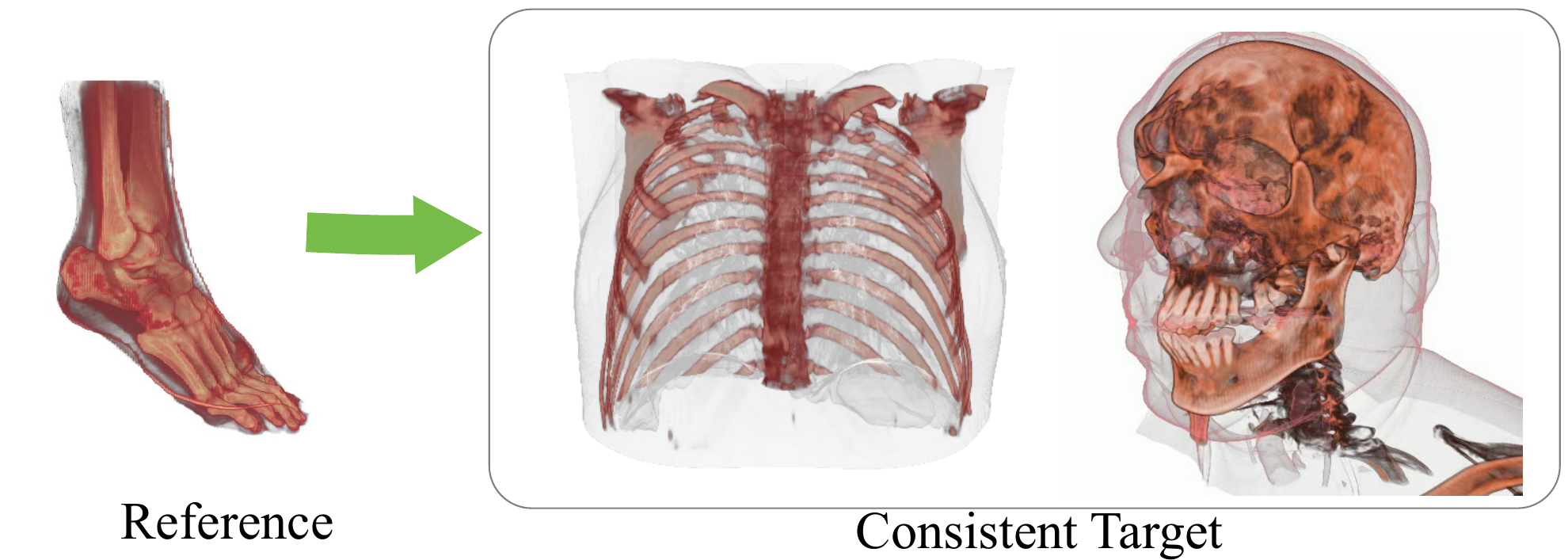}
    \caption{Image-guided TF design for consistent visualization. Using a rendering from a foot volume as a reference, our system generalizes the visualization standard to other anatomical datasets (chest and head).}
    \label{fig:case2}
    \vspace{-1em}
\end{figure}

\begin{figure*}[!t]
    \centering
    \includegraphics[width=\textwidth]{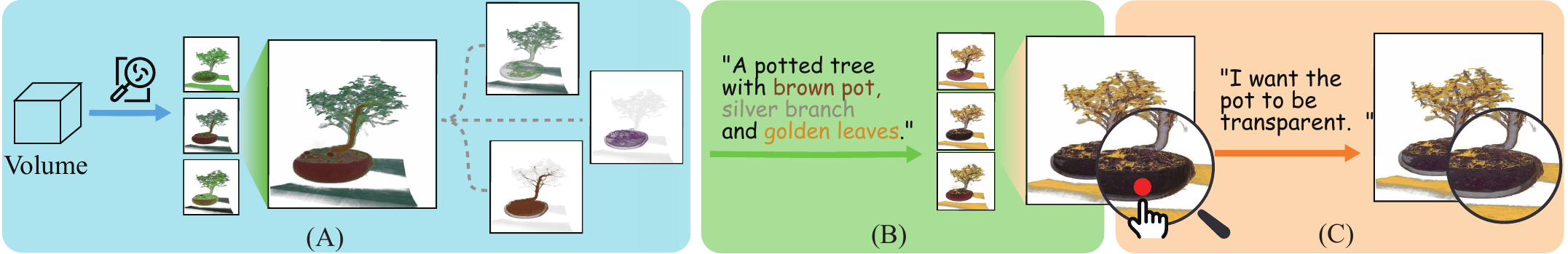}
    \caption{The interactive transfer function design process for the first case in Sec. ~\ref{sec:case_study}. The user (A) develops understanding of the volume through initialization results, (B) customizes the visualization using natural language commands, and (C) refines the final rendering by selecting and implementing specific modifications.}
    \label{fig:case1}
    \vspace{-1em}
\end{figure*}
\section{Case Studies}\label{sec:case_study}

We demonstrate the versatility and utility of our approach through three detailed case studies that collectively showcase our system's capabilities across various visualization needs. These cases illustrates how our system enables intuitive transfer function exploration and design through simple natural language descriptions or reference images.

\subsection{Image-Guided TF Design for Consistent Medical Visualization}
\label{sec:case_med}

This case study demonstrates how our image-guided transfer function design enables convenient deployment of consistent visualization standards across different anatomical datasets (Fig. \ref{fig:case2}). In medical imaging, radiologists often need to analyze scans obtained from diverse sources, such as different devices or patients. However, variations in individual anatomy and scanning parameters introduce distributional shifts, making it difficult for a transfer function carefully designed for one scan to achieve the same visual effect when directly applied to another. Our approach addresses this issue by ensuring that all datasets are visualized under a unified standard.

In this case study, a radiologist first used our system to design a satisfactory visualization scheme for a left foot CT dataset. This scheme clearly displayed different tissue types including bones, muscles, and skin, and the radiologist decided to use it as the visualization standard for further analyses. Subsequently, the radiologist acquired two new CT datasets, which our system's automatic exploration identified as chest and head volumes respectively.

The radiologist aimed to apply the validated visualization standard to these two new target datasets to achieve visual consistency. Therefore, the user input the previous left foot rendering as the reference image to the system. The system subsequently generated multiple candidate transfer functions for both the chest and head datasets. The user selects the result of interest, and confirmed the outcomes through rotation and scaling in the volume interaction view (Fig. \ref{fig:interface}, \raisebox{-0.3em}{\includegraphics[height=1.2em]{figs/c.pdf}}). The results demonstrated consistent visual effects in all structural elements across different datasets: all volume data bones (foot bones, ribs, and skull bones) displayed identical opacity and color tones, while muscle and skin tissues maintained semi-transparent effects. By successfully deploying consistent visualization standards across different anatomical regions, the radiologist could rapidly analyze chest and head structures using familiar visual cues within a unified visual framework. 

This cross-data-distribution visual consistency not only reduces cognitive load when switching between different datasets, but also enhances standardization and reproducibility in volume data visualization. (F), (G) in Fig. \ref{fig:teaser} further demonstrate our image guidance capability in reproduction, where we used images from other papers as visual standards to guide volume data visualization. Even without knowledge of the original rendering parameters (incl. TF, camera poses), our rendering results achieve consistent visual standards with the reference images, accomplishing feature-level alignment.

\subsection{Interactive TF Exploration and Customization}

Our second case study demonstrates how our interactive system facilitate text-guided visualization design for a botanical volume data (Fig. \ref{fig:case1}).

During the initial exploration phase using the feature selection view (Fig. \ref{fig:interface}, \raisebox{-0.3em}{\includegraphics[height=1.2em]{figs/d.pdf}}), the user discovered that the dataset represented a detailed potted plant model with distinct structural components. After identifying the pot, branches, and leaves, the user wanted to visualize these elements with clear visual differentiation.

The user formulated their visualization intent through natural language: ``A potted tree with brown pot, silver branch and golden leaves.'' This prompted the system to generate several candidate visualizations displayed in the recommendation gallery view (Fig. \ref{fig:interface}, \raisebox{-0.3em}{\includegraphics[height=1.2em]{figs/b.pdf}}). While reviewing these options, the user gained a clearer understanding of the dataset's structure but noticed a significant insight—the completely opaque pot was concealing potentially important root structures within.

Based on this discovery, the user decided to refine the visualization to expose these hidden internal features. Using the volume interaction view (Fig. \ref{fig:interface}, \raisebox{-0.3em}{\includegraphics[height=1.2em]{figs/c.pdf}}), they specifically selected the pot region by simply clicking  and requested to make it ``transparent.'' This targeted modification revealed previously obscured root structures while preserving the distinctive appearance of the branches and leaves.

The resulting visualization offered new insights into how the root system connected with the branches within the pot—spatial relationships that weren't visible in the previous rendering. The user could now observe the complete plant structure from root to leaf, providing a comprehensive understanding of the model's internal organization. Satisfied with how this visualization revealed both the component differentiation and their interconnections, the user saved the transfer function and final rendering for future analysis.

\subsection{Sketch-based TF Design for Visual Intent Expression}

Our third case study demonstrates how non-expert users can effectively customize medical volume visualization through intuitive sketch-based interaction. This case involves a non-medical user who obtained personal dental CT data and sought to understand tooth structure through our visualization system (Fig. \ref{fig:teaser}, (H)).

The user began with CT scan data of their tooth. During the initial exploration phase, our system automatically analyzed the volume data and generated several candidate transfer functions. The user reviewed these system-recommended renderings and identified the tooth structure, observing the crown, root, and other major components. 

The user sought to enhance visual contrast by applying different colors to distinct tooth regions. Lacking precise anatomical terminology, the user could not verbally articulate these specific regions. Instead of using text input, the user sketched directly on the rendering, marking the crown top in red, the crown middle in yellow, and the root in blue. The system received this sketched image and interpreted it as "red crown, yellow dentin, blue root" then optimized the transfer function accordingly. The user found the resulting color differentiation improved structural detail examination and adopted this rendering configuration for subsequent dental analysis.

\section{Experiments and Results}\label{sec:experiment}
This section presents both the design and outcomes of two controlled experiments evaluating our system. First, we validate our MLLM-guided evaluator by comparing its assessments with expert opinions. Second, we analyze the influence of evolutionary algorithm parameters on transfer function optimization performance, identifying optimal parameter configurations for different computational constraints. Last, we discuss the computational cost of the entire process.

\begin{figure*}[!t]
    \centering
    \includegraphics[width=\textwidth]{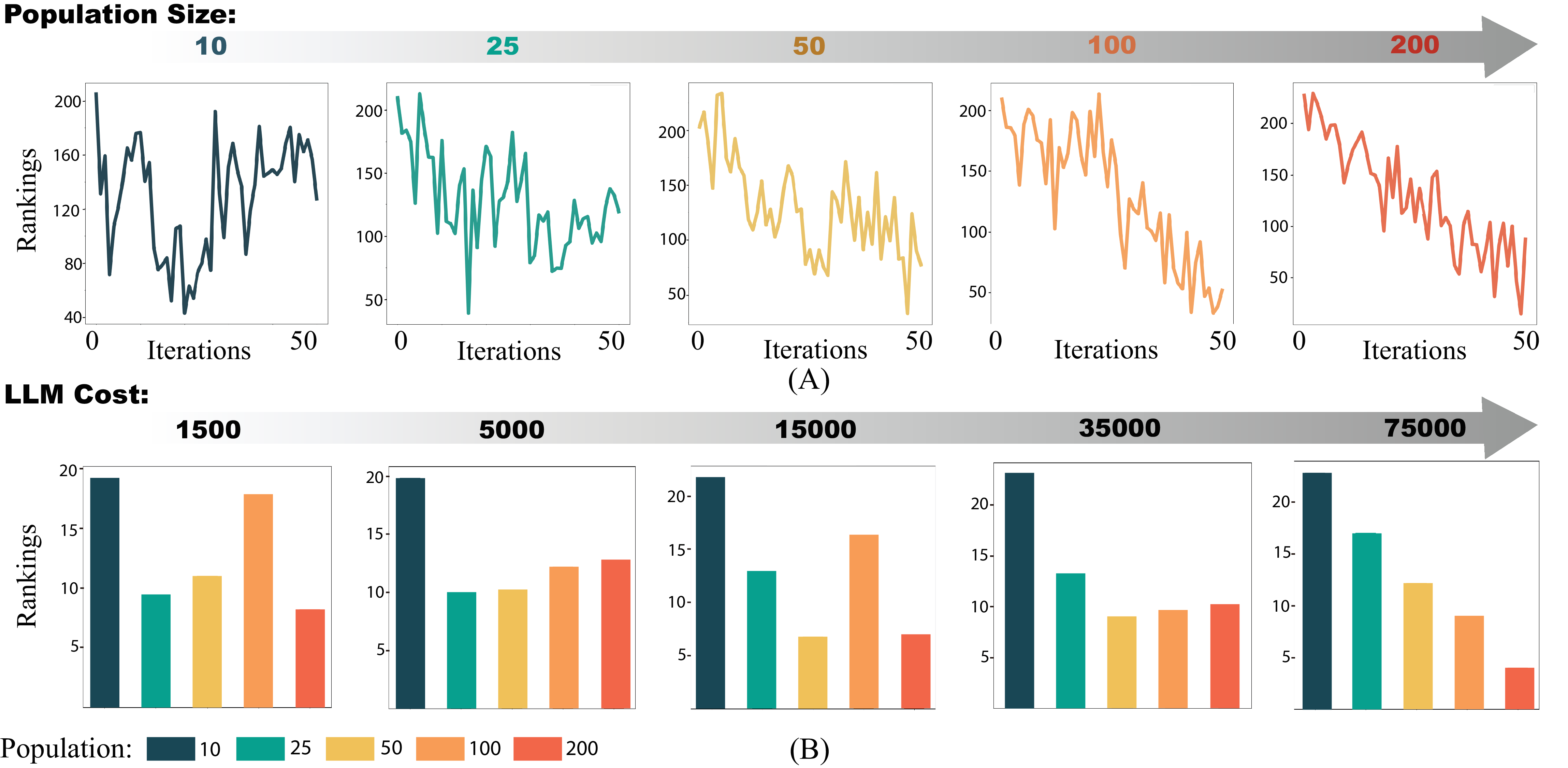}
    \caption{Analysis of evolutionary algorithm parameters on transfer function optimization. (A) Average rankings over iterations for different population sizes (10-200), showing that larger populations achieve more stable improvements with continued iterations while smaller populations show early rapid gains but unstable convergence. (B) Comparison of optimization effectiveness across different computational budgets (measured by LLM API calls), demonstrating that medium-small populations (25-50) perform better under limited resources, while larger populations (100-200) achieve superior results when computational resources are abundant.}
    \label{fig:experiments}
    \vspace{-1em}
\end{figure*}

\subsection{Human-Aligned Evaluator Evaluation} \label{sec:evaluator_exp}
To systematically assess the reliability of our automatic evaluator, we designed a human-machine comparison experiment using medical volume data. The medical domain was selected because of its well-established quality assessment criteria and the availability of domain experts. Four representative medical volume datasets: feet, head, chest, and tooth, were used as test cases. For each volume dataset, we prepared 20 rendered images with quality uniformly distributed from high to low. These images were paired to form 190 comparison samples. We recruited 28 medical domain experts through online platforms, all of whom were either practicing physicians or medical students in clinical training. Each participant was instructed to compare all 190 image pairs and select their preferred image in each pair, with the option to declare a tie when differences were negligible. All participants completed the assessment within 30 minutes and received approximately \$7 USD as compensation. To establish ground truth labels from multiple expert annotations, we applied majority voting: for each image pair, the preference selected by the majority of experts was taken as the consensus judgment.

Our algorithmic evaluation methods were subjected to identical pair-wise comparison tasks to enable direct comparison with human judgments. To evaluate the consistency between our method and human judgments, we adopt the evaluation framework proposed by GPTEval3D \cite{gpteval3d}. For each comparison task, we model both human and algorithmic evaluations as Bernoulli distributions, with probabilities $q_i$ (human) and $p_i$ (our method) of selecting the first image. We quantify the agreement score using $\frac{1}{N}\sum_{i=1}^{N} p_iq_i + (1-p_i)(1-q_i)$. This metric yields values in the range [0,1], where 1 indicates perfect agreement (both consistently prefer or reject the same renders) and 0 indicates complete disagreement. Higher values indicate stronger consistency between human and algorithmic preferences.

Within this evaluation framework, we conducted two complementary experiments to comprehensively validate the robustness of our MLLM-guided evaluator. The first experiment examines cross-dataset and cross-model generalizability, while the second validates our designed evaluation aspects.

\textbf{Cross-dataset and cross-model analysis.} Table \ref{tab:dataset_comparison} compares the performance of our MLLM-guided evaluator (using both GPT-4o and Gemini-2.0 Flash) against the information entropy baseline across four anatomical datasets. The results reveal significant variability in information entropy's performance across different datasets: it achieves reasonable agreement scores for some datasets (e.g., 0.66 for CHEST) but performs poorly on others (e.g., 0.36 for FEET). This inconsistency suggests that traditional quantitative metrics struggle to adapt to varying dataset complexities and structural characteristics. In contrast, our MLLM-guided evaluators demonstrate consistently superior and stable performance across all datasets. The GPT-4o version achieves agreement scores ranging from 0.77 to 0.85, while the Gemini-2.0 Flash version maintains scores between 0.70 to 0.76. Notably, both MLLM variants significantly outperform the information entropy baseline on every dataset, with improvements ranging from 0.09 to 0.41 points.

\begin{table}[!t]
\centering
\caption{Performance improvement of our evaluator over GPT-4o and Gemini-2.0 Flash across anatomical datasets.}
\label{tab:dataset_comparison}
\begin{tabular}{c@{\hspace{12pt}}c@{\hspace{12pt}}c@{\hspace{12pt}}c}
\toprule
\textbf{Datasets} & 
\makecell[c]{\textbf{IntuiTF} \\ \textbf{(GPT-4o)}} & 
\makecell[c]{\textbf{IntuiTF} \\ \textbf{(Gemini-2.0} \\ \textbf{Flash)}} & 
\makecell[c]{\textbf{Information} \\ \textbf{Entropy} \\ \textbf{(baseline)}} \\
\midrule
HEAD & 0.7711 & 0.7026 & 0.5346 \\
CHEST & 0.8480 & 0.7573 & 0.6604 \\
FEET & 0.7868 & 0.7211 & 0.3580 \\
TEETH & 0.7661 & 0.7515 & 0.4933 \\
\bottomrule
\end{tabular}
\end{table}

These results demonstrate two key advantages of our approach: (1) our evaluator consistently outperforms traditional quantitative metrics across different datasets, demonstrating strong generalization capability, and (2) our evaluator effectively generalizes across different state-of-the-art multimodal models, confirming the robustness and transferability of our design. The consistent performance across both MLLMs and all datasets indicates that our evaluator provides more comprehensive and human-aligned assessment for direct volume rendering quality.

\textbf{Evaluation aspect validation.} To further validate the effectiveness of our designed evaluation aspects, we conducted an analysis across the three quality aspects used in our evaluator: information richness (IR), feature discrimination (FD), and color-content harmony (CH). Table \ref{tab:aspect_comparison} shows agreement scores between expert evaluations and our evaluator implemented with different MLLM backends (GPT-4o and Gemini-2.0 Flash) for each individual aspect, with the rightmost column presenting the combined agreement when all three aspects are evaluated together. All scores are averaged across our four volumetric datasets. The findings indicate that information richness, feature discrimination, and color-content harmony achieved reasonable agreement scores, suggesting these aspects align well with human visual perception in medical volume rendering contexts. Moreover, both MLLM-guided evaluators (GPT-4o and Gemini-2.0 Flash) performed consistently well, with GPT-4o demonstrating slightly superior performance.

\begin{table}[!t]
\centering
\caption{
Agreement scores of our evaluator equipped with two widely used MLLMs, i.e. GPT-4o and Gemini-2.0 Flash, on medical volume rendering dimensions (higher indicates better performance). We measure the agreement score across different visual quality aspects: information richness (IR), feature discrimination (FD), and color-content harmony (CH). The IR\&FD\&CH column represents agreement when all three aspects are considered jointly in the assessment.}

\centering
\label{tab:aspect_comparison}
\begin{tabular}{c|cccc}
\toprule
\textbf{Models} & \multicolumn{4}{c}{\textbf{Aspects}} \\
\cmidrule{2-5}
 & \textbf{IR} & \textbf{FD} & \textbf{CH} & \textbf{IR\&FD\&CH} \\
\midrule
GPT-4o & 0.7872 & 0.7930 & 0.7192 & 0.7804 \\
Gemini-2.0 Flash & 0.7413 & 0.7331 & 0.6482 & 0.7419 \\
\bottomrule
\end{tabular}
\vspace{-1em}
\end{table}

\subsection{Evolution-Driven Explorer Evaluation}

In this experiment, we employed the three aspects validated in Sec. \ref{sec:evaluator_exp} as evaluation criteria to analyze how two critical evolutionary algorithm parameters, population size and iteration count, influence transfer function optimization performance. Since our MLLM evaluator uses comparative assessment rather than generating absolute quantitative scores, we utilize the Elo rating method described in Sec. ~\ref{subsec:evaluator} for evaluation.
To compare the impact of different hyperparameters on optimization results, we follow this process: For each hyperparameter configuration, we select the top $10$ individuals from its final population as quality representatives. We then combine all these representatives across different configurations into a collective pool for comprehensive comparison. Using our Elo rating method, we calculate ratings and rankings for each individual in this pool. Finally, we assess the quality of each configuration by computing the average ranking of its representatives, with lower average rankings indicating superior optimization outcomes.

Fig. \ref{fig:experiments}(A) illustrates the relationship between iteration count and average ranking across five different population sizes ($10$, $25$, $50$, $100$, and $200$). Our observations reveal distinct patterns: small populations ($10$ and $25$) improve rapidly in early stages but converge soon after with unstable rankings, medium populations ($50$) show significant improvement before stabilizing around $30$ iterations, and large populations ($100$ and $200$) continue to improve even after $50$ iterations with steadily decreasing average rankings.

To further analyze computational efficiency across configurations, we compared different population sizes with similar MLLM api call budgets. Considering the $n\log_{2}(n)$ complexity of Elo rating calculations, different population sizes require different computational resources per iteration. (e.g., one iteration's API budget with a population of $200$ is similar to $50$ iterations with a population of $10$.). Using the computational cost of $50$ iterations for each population size as benchmarks, we selected equivalent resource nodes ranging from $10 \times 50 \times \log_{2}(10)$ to $200 \times 50 \times \log_{2}(200)$ for comparison. Fig. \ref{fig:experiments} (B) demonstrates that under limited computational resources, medium-small populations ($25$-$50$ individuals) obtains the best optimization results; however, as computational resources increase, larger populations progressively demonstrate quality advantages, with ranking curves showing a completely downward trend at the maximum resource node. This indicates that abundant resources enable larger populations to consistently produce higher-quality results.

Based on this analysis, we propose the following practical optimization guidelines: under constrained computational resources, medium-small populations ($25$-$50$ individuals) with moderate iterations ($20$-$30$) provide the most efficient improvements; with abundant resources, larger populations ($100$-$200$ individuals) can achieve continuous quality enhancements.

\subsection{Computational Cost Analysis}

Our method's computational procedure consists of three main components: (1) volume rendering for each candidate, (2) crossover \& mutation operations (Sec.~\ref{subsec:evolution}), and (3) fitness evaluation (Sec.~\ref{subsec:evaluator}). Both rendering and crossover \& mutation operations are fully parallelized, resulting in constant-time overhead of $O(1)$ per iteration.

The dominant cost arises from fitness evaluation, which requires querying a large language model to compute fitness scores. According to the Elo rating strategy (Sec.~\ref{subsec:evaluator}), each individual participates in $\log_2(m) + 2$ comparison rounds, where $m$ represents the population size. Since all comparisons within a round execute in parallel, the effective cost per iteration is $(\log_2(m) + 2) \cdot T$, where $T$ denotes the average API response time (treated as a constant). Consequently, the total runtime over n iterations is about $n \log_2 m \cdot T$, corresponding to a time complexity of $O(n \log_{2} m)$.

Empirical observations show that each iteration takes approximately 90 seconds for population of $50$, with most time attributed to external API communication rather than local computation. The cost for one round of LLM API calls is approximately \$0.23 USD using Gemini 2.0 Flash as an example. In practice, the transfer function quality becomes acceptable within 6 iterations and stabilizes around 10-15 iterations. While the current computational overhead is manageable for practical applications, future improvements could leverage faster APIs or more efficient model architectures to reduce both time and cost.

\section{Expert Interviews}
\label{sec:expert}

We conducted a set of expert interviews to obtain qualitative feedback on the usability, potential, and limitations of IntuiTF. We recruited four domain experts (E1–E4) from local universities and industry. E1 and E2 are researchers in scientific visualization, E3 is a researcher in computer vision and large language models, and E4 is a professional physician from a local hospital. At the beginning of each session, we introduced the background and demonstrated how to use the system. Experts were then asked to freely explore different volumetric datasets with IntuiTF. After the exploration phase, we conducted semi-structured interviews focusing on three aspects: (1) overall impressions of the system, (2) perceived strengths and limitations, and (3) possible future directions.

All experts expressed generally positive impressions of the system, agreeing that IntuiTF produces high-quality results that are both visually compelling and informative. E3 highlighted that the system is simple and accessible even for non-specialists.

Several experts emphasized specific strengths of the system. E2 noted that the initial renderings generated by IntuiTF clearly conveyed volumetric structures, and that subsequent customizations aligned well with the given visualization intents. E1 pointed out that designing transfer functions is often difficult to articulate in words, and reckoned the ability to use reference images as a design language is highly valuable.

Experts considered the image-guided transfer function design particularly beneficial in real-world scenarios. E4 specifically highlighted this capability's value in medical volume visualization, where consistent transfer function design across structurally similar datasets facilitates comparative analysis (see Sec~\ref{sec:case_med}).

Experts also identified several areas for improvement. First, the transfer function optimization still requires considerable waiting time. E3 suggested that the system could benefit from incorporating prior knowledge from exploring similar volumetric datasets. Second, E2 noted that for abstract datasets lacking semantic information (e.g., simulation data), both textual and image-based descriptions struggle to capture visualization requirements effectively. Users can only describe color preferences or provide stylistic reference images, making it challenging to achieve complete alignment with their visualization needs.
\section{Discussion}

In this paper, we propose a transfer function optimization approach that bridges the semantic gap between high-level user intents and low-level transfer function parameters. Here, we discuss some future directions for future research to enhance our approach.

\textbf{Knowledge distillation of exploration results.} While our current approach demonstrates effectiveness in bridging semantic intents with rendering parameters, its reliance on multiple MLLM API calls introduces additional computational costs. A promising direction for addressing this limitation is to distill knowledge from successful exploration results into a specialized generative model for volume rendering. Such a model would learn from a curated dataset of semantic descriptions paired with their corresponding optimized transfer functions, effectively capturing the underlying mapping patterns discovered through our framework. Beyond computational efficiency, this distillation process could potentially reveal interpretable patterns in how semantic concepts map to rendering parameters, providing insights into the relationship between natural language descriptions and effective visualization strategies. The distilled model could also enable more nuanced parameter generation by learning from the variations in transfer functions that achieve similar semantic goals across different datasets. This would create a more scalable and responsive system while maintaining the semantic understanding capabilities of the original approach.

\textbf{Incorporating image generation models for post-processing refinement.} Leveraging image generation models could enhance rendered results through sophisticated post-processing refinements such as contrast improvement, noise reduction, or stylization while preserving structural information. By conditioning on both volume rendering output and semantic descriptions, this integration could achieve more coherent results that better match domain experts' mental models and enable iterative refinement through natural language feedback.

\section{Conclusion}

We propose a framework, IntuiTF, an intuitve transfer function design framework. We formulate the expert design process into "Trail-Insight-Replanning" cycle, and implement the cycle through an evolution-driven explorer and an MLLM-guided human-aligned evaluator. 
Based on this approach, we further develop an interactive transfer function design system and use three case studies to demonstrate various applications of our system. Besides, we show the effectiveness of our approach through extensive experiments.

This work illustrates how MLLMs can serve as perceptual evaluators in transfer function design. We believe this framework opens promising avenues for future research in scientific visualization.


\bibliographystyle{abbrv-doi-hyperref}

\end{document}